\documentclass[aps,pra,reprint,showpacs,floatfix,twocolumn]{revtex4-2}
\usepackage{amsmath,amsthm,amsfonts,amssymb,amscd}
\usepackage[utf8]{inputenc}
\usepackage[english]{babel}
\usepackage{babel}
\usepackage{graphics}
\usepackage{graphicx}
\usepackage{textcomp}
\usepackage{latexsym}
\usepackage{enumerate}
\usepackage[caption=false]{subfig}
\usepackage[colorinlistoftodos, portuguese, textwidth=140pt]{todonotes}
\usepackage{hyperref}

\begin{document}
\title{Gouy phase effects in the frequency combs of an optical parametric oscillator}

\author{R. F. Barros, G. B. Alves, A. Z. Khoury}
\affiliation{Instituto de F\'isica, Universidade Federal Fluminense, CEP 24210-346, Niter\'oi-RJ, Brazil}
\email{rafael.fprb@gmail.com}

\begin{abstract}
We investigate the transverse mode effects on the frequency combs generated by a type-I optical parametric oscillator (OPO) below threshold. We take the diffraction effects fully into account, considering usual OPO architectures instead of the self-imaging design. Further, we show that an OPO pumped with structured light produces multiple frequency combs simultaneously, corresponding to different combinations of spatial modes in the downconverted fields. This result may apply to the production of hybrid multipartite entanglement in the spatiotemporal modes of a single OPO.


\end{abstract}
\maketitle

\section{Introduction}

Optical parametric oscillators (OPO) are the quintessential sources of non-classical states of light. The simplest examples are the single-mode squeezed states produced by a degenerate OPO below threshold \cite{Wu:86,Wu:87}, and the pairwise Einstein-Podolski-Rosen (EPR) correlations produced in the non-degenerate case \cite{Ou:92,Drummond:90, Chalopin:12}. More recently, different schemes using the OPO for the generation of multipartite entanglement have been successfully implemented. In the spatial domain, for example, an OPO pumped with structured light was shown to produce quadripartite entanglement between transverse modes of the downconverted fields \cite{Liu:16,Cai:18}. In the spectral domain, a dual-frequency pump was used to entangle $60$ modes in the quantum optical frequency comb of a single OPO \cite{Chen:14}. It is also worthwhile to mention recent developments in the production of multipartite entanglement with synchronously pumped OPOs \cite{Roslund2014,Cai2017}.

Generating such highly multipartite entangled states is of utmost importance for continuous-variable quantum information \cite{Braunstein:05}, especially for the growing field of measurement-based quantum computing \cite{Briegel:01,Gu:09}. In this model, a highly entangled substrate -- a cluster state -- is prepared, and adaptive measurements are performed to implement algorithms. A standard example of a universal cluster state is a 2D square lattice, which can be produced in the OPO frequency comb with different approaches  \cite{Menicucci:08, Zhu:20}.

In the spatial domain, it has been proposed that a large-scale cluster state can be produced in the OPO using a structured pump to entangle a comb of spatial modes carrying orbital angular momentum (OAM)\cite{Zhang:17}. A hybrid approach has been recently explored, in which the spatial and spectral modes of the OPO are used to generate spatiotemporal graph states \cite{Yang:20}. In both cases, however, the diffraction effects on the resonator modes were neglected by considering a specially designed self-imaging cavity \cite{Lopez:09}. The transverse mode effects on the frequency combs of a generic OPO are still unclear.

This work aims to provide a detailed study on the spatially structured frequency combs produced by a type-I OPO below threshold. We calculate the diffraction-affected longitudinal modes, considering arbitrary spatial structures in the pump and the downconverted fields, as well as different OPO architectures. The Gouy phase of the different transverse modes is explicitly taken into account in order to correctly derive the resonance conditions. The Gouy phase effects in nonlinear optical processes have been extensively investigated in several recent works \cite{Pereira:17,Wu:19,Pires:19,Pires:20,Wu:20,Buono:20, Offer:20}.  Here, we show that these effects play a crucial role in the spatially structured frequency combs produced by an OPO. In this way, we extend the previous works in the literature by addressing the comb generation in more realistic experimental conditions. 

The paper is organized as follows. In the next section, we introduce the basic OPO equations and the evolution of the Gouy phase in a cavity round-trip. In Section \ref{Sec-longitudinal-modes} we study the transverse mode effects on the OPO longitudinal modes. We discuss the spatially structured frequency combs in Section \ref{Sec-frequency-combs}, and their change due to reflection phase shifts in Section \ref{Sec-reflection-phases}. The OPO spectrum for alternative OPO architectures is studied in Section \ref{Appendix-alternative-cavities}.  In Section \ref{conclusion} we draw our conclusions.  Appendix \ref{Appendix-bandwidth} provides a discussion on the bandwidth limitation to the frequency comb generation.

\section{Basic equations}
An optical parametric oscillator consists of a nonlinear crystal embedded in an optical cavity. When pumped by an external field $E_0$ of frequency $\nu_0$, it generates $E_1\,$(signal) and $E_2\,$(idler) with frequencies $\nu_1$ and $\nu_2\,$, respectively. While the energy conservation imposes $\nu_1+\nu_2=\nu_0$ in the nonlinear process, the frequency difference between signal and idler, i.e., the beat-note
\begin{equation}
 \Delta \nu=\nu_1-\nu_2\,,
 \label{beat-note-def}
\end{equation}
is only restricted by the cavity resonances. The allowed values of the beat-note define the longitudinal modes of the OPO. 

Let us express the time-dependent electric fields as
\begin{equation}
\mathbf{E_j}(\mathbf{r},t)=Re\{\mathbf{E_j}(\mathbf{r})e^{-i\omega_j t}\}\,,\qquad j=0,1,2\, ,
\end{equation}
where $\omega_j=2\pi\nu_j$, and the complex amplitudes $\mathbf{E_j}(\mathbf{r})=\mathbf{E}(\mathbf{r},\omega_j)$ contain the field spatial structure and polarization. If $z$ is the propagation direction, they can be expanded as
\begin{equation}
\mathbf{E_j}(\mathbf{r})= \sum_{mn}u_{mn}^j(\mathbf{r})\alpha^j_{mn}(z)e^{ik_jz}\boldsymbol{\hat{\epsilon}}_j\,,
\label{E-field}
\end{equation}
where $k_j={2\pi n_j \nu_j}/{c}$ is the wave vector, $n_j$ is the refractive index and $c$ is the speed of light in vacuum. The functions $\alpha^j_{mn}(z)$ are the amplitudes of the transverse modes $u^j_{mn}(\mathbf{r})$. They are assumed to be slowly varying functions of $z$, which only change appreciably over distances much larger than the wavelength $\lambda_j=2\pi/ k_j$. 

Assuming paraxial propagation, the modes $u^j_{mn}(\mathbf{r})$ are solutions of the so-called paraxial wave equation
\begin{equation}
\nabla_\perp^2 u^j_{mn}+2k_j\frac{\partial u^j_{mn}}{\partial z}=0\,,
\label{paraxial}
\end{equation}
where $\nabla_\perp \equiv \partial^2_x+ \partial^2_y $ is the transverse Laplacian. They generate different families of solutions, such as the well-known Hermite-Gaussian (HG) and Laguerre-Gaussian (LG) modes. Here we focus only on the HG modes, which are best suited to describe astigmatic systems such as ring cavities and birefringent crystals. 

\begin{figure}
\includegraphics[scale=0.37]{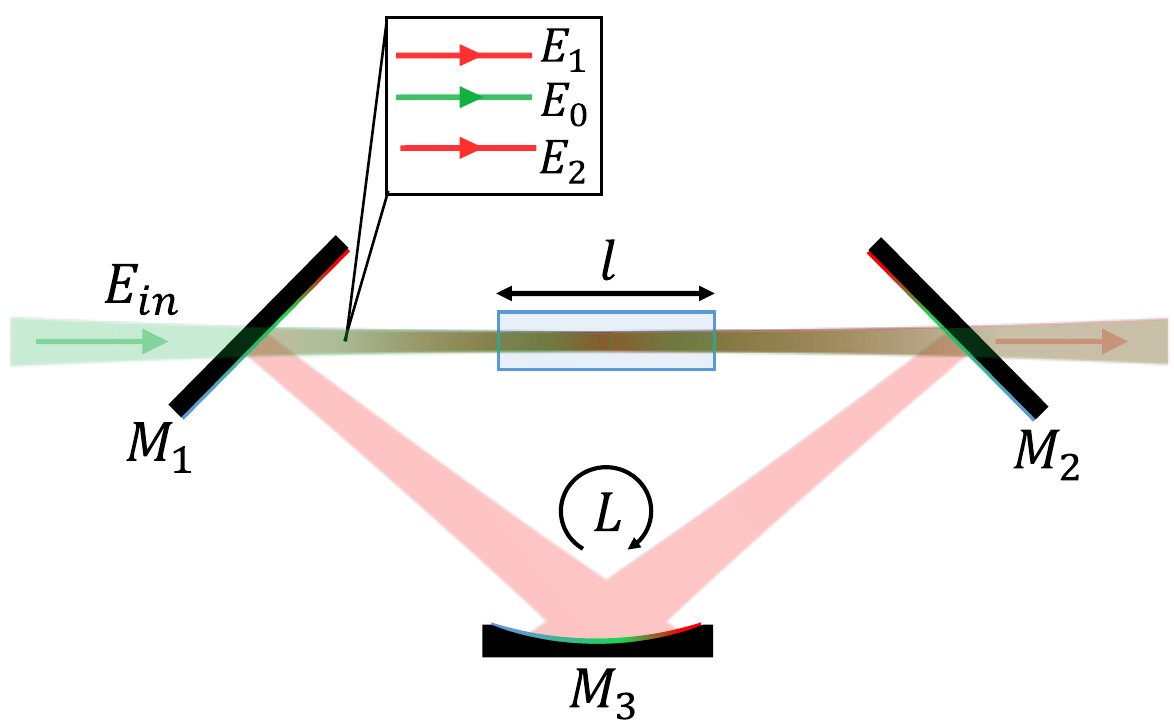}
\caption{Optical parametric oscillator with a symmetric triangular ring cavity. Mirrors $M_1$ and $M_2$ are flat, while $M_3$ is concave with radius $R$. The geometrical length of the round-trip is $L$, and the crystal length is $l\,$.}
\label{FIG-OPO}
\end{figure}

The Hermite-Gaussian mode functions are
\begin{eqnarray}
\!\!\!\!u^j_{mn}(\mathbf{r})&=&C_{mn}^j\,H_m\left(\frac{\sqrt{2} x}{w_j(z)}\right) H_n\left(\frac{\sqrt{2} y}{w_j(z)}\right)\\
&\times&\exp\left[-\frac{x^2+y^2}{w_j^2(z)} +i\frac{k_j(x^2+y^2)}{2R_c(z)} +i\phi^j_{mn}(\mathbf{r})\right]\, ,\nonumber
\label{mode}
\end{eqnarray}
where $C^j_{mn}$ is a normalization constant and $H_n\left(x\right)$ is the $n$th-order Hermite polynomial. The argument $\phi^j_{mn}(\mathbf{r})$ is the Gouy phase, defined as
\begin{equation}
\phi^j_{mn}= -(S_j+1)\arctan\left(\frac{z}{z_0^{(j)}}\right)\,,
\end{equation}
where $S_j=m_j+n_j$ is the order of the transverse mode. The parameters  $w_j(z)$,  $z_0^{(j)}$ and $R_c^{(j)}(z)$ are the beam diameter, its Rayleigh length and the curvature radius of its wavefront, respectively. They are given by
\begin{equation}
\begin{aligned}
w_j(z)&=w_j(0)\sqrt{1+(z/z_0^{(j)})^2}\,,\\
R_c^{(j)}(z)&=z[1+(z_0^{(j)}/z)^2]\,,\\
z_0^{(j)}&=\pi w_j^2(0)/\lambda_j\,.
\end{aligned}
\end{equation}
where $w_j(0)$ is the beam waist.

\subsection{Phase evolution in a round trip}

Consider the OPO illustrated in Fig.\ref{FIG-OPO}, with a triangular ring cavity formed by two flat mirrors and one concave mirror of radius $R$. Apart from constant phase factors added by the cavity mirrors, that will be discussed in Section \ref{Sec-reflection-phases}, the phase accumulated in each round trip is given by
\begin{equation}
\phi_ j=\frac{2\pi\nu_j}{c}L_j-\phi_G^j\,.
\label{delta-phi}
\end{equation}
where
\begin{equation}
\phi_G^j=2(S_j+1)\arctan\left(\frac{2z_0^{(j)}}{2R-D_j}\right)\,,
\label{Gouy-phase-ring}
\end{equation}
is the accumulated Gouy phase. The cavity is assumed to be mode-matched, in which case $z_0^{(j)}=\sqrt{D_j(2R-D_j)}/2$. The quantities $L_j$ and $D_j$ are the optical path length and the effective diffraction length of a round trip, respectively, and are given by 
\begin{eqnarray}
L_j&=&L+l(n_j-1)\,,\label{optical-path-length}\\
D_j&=&L+l\left(\frac{1}{n_j}-1\right)\label{diffraction-length}\,,
\end{eqnarray}
where $L$ is the geometrical length of the round-trip and $l$ is the crystal length. It will show convenient to define the quantities $\bar{L}$, $\bar{D}$ and $\bar{z}_0$, which are evaluated at the frequency degeneracy condition $n_1=n_2=n(\nu_0/2)$.

The condition for perfect resonance is that the field interfere constructively after each round-trip, which occurs when $\phi_j=2p_j\pi\,, p_j\in\mathbb{Z}\,$. The phase shift with respect to the nearest resonance is the detuning $\delta\phi_j$, defined as
\begin{equation}
\delta\phi_j=\phi_j-2p_j\pi\,.
\label{Detuning}
\end{equation}
The detuning is related to the geometrical displacement from resonance $\Delta L_j=L_j-L_j^{res}$ as
\begin{equation}
\frac{\delta\phi_j}{2\pi}=\frac{\Delta L_j}{\lambda_j}\,.
\label{detuning-displacement}
\end{equation}
We will refer to equation \eqref{Gouy-phase-ring} as the accumulated Gouy phase. This expression will be adapted to different cavity designs in Section \ref{Appendix-alternative-cavities}.

\subsection{OPO equations}

Let us consider the slowly-varying amplitudes $\alpha_j=\alpha^j_{m_j,n_j}(z)$ of the pump, signal and idler fields ($j=0,1,2$, respectively) for a given set of transverse modes. Neglecting absorption losses inside the crystal, the evolution of the fields in the parametric interaction is described by the following equations \cite{Eckardt:91}
\begin{eqnarray}
\frac{d\alpha_0}{dz}=ig\alpha_1\alpha_2 \,,\label{stat-OPO0}\\  
\frac{d\alpha_1}{dz}=ig\alpha_0\alpha_2^*\,,\label{stat-OPO1}\\
\frac{d\alpha_2}{dz}=ig\alpha_0\alpha_1^* \,.\label{stat-OPO2}
\end{eqnarray} 
The non-linear coupling coefficient $g$ is given by
\begin{equation}
g=\chi^{(2)}\int_{-l/2}^{l/2} dz \Lambda^{m_0m_1m_2}_{n_0n_1n_2}(z)e^{i\Delta k z}\,,
\label{nonlinear-coupling}
\end{equation}
where  $\chi^{(2)}$ is proportional to the second-order nonlinear susceptibility and $l$ is the crystal length. Also, $\Delta k=k_0-k_1-k_2$ is the wave vector mismatch and 
\begin{equation}
 \Lambda^{m_0m_1m_2}_{n_0n_1n_2}(z)=\int d^2r\,u^0_{m_0n_0}(\mathbf{r})u^{1*}_{m_1n_1}(\mathbf{r})u^{2*}_{m_2n_2}(\mathbf{r})\,,
\end{equation}
is a transverse overlap integral. These integrals originate a number of selection rules, which were extensively discussed in Refs.\cite{Schwob1998, Alves2018,Pereira:17}

We consider a doubly-resonant cavity, which is transparent at the pump wavelength and has low losses for the downconverted fields. A nontrivial stationary solution for the fields is obtained when the parametric interaction compensates the amplitude and phase changes in the cavity round trip, which implies
\begin{eqnarray}
\delta\phi_1&=&\delta\phi_2=\delta\phi \label{equal-detunings}\\
|\alpha_0|^2&\geq&\frac{1}{|g|^2}(\delta\phi^2 +\pi^2/F^2)\label{DROPO-threshold}\,,
\end{eqnarray}
where $F$ is the cavity Finesse for the downconverted fields, and the RHS of \eqref{DROPO-threshold} defines the OPO threshold.

Even when \eqref{DROPO-threshold} is satisfied for many modes, the OPO produces only a single pair of bright beams that survives the competition for the nonlinear gain \cite{Alves2018}. Here we are particularly interested in the below-threshold operation, in which the OPO produces a large set of EPR pairs with correlations that increase as the pump approaches the threshold \cite{Drummond:90}. In the next section we analyze the influence of diffraction and transverse mode structure on the longitudinal modes resonances ($\delta \phi=0$), where the threshold is minimum.




%


\section{Diffraction-affected longitudinal modes}\label{Sec-longitudinal-modes}

 The longitudinal modes of the OPO were deeply investigated by different authors \cite{debuisschert_type-II_1993,Eckardt:91} in the plane wave approximation. However, little attention has been given to the inclusion of diffraction and spatially structured fields. This section is divided in two parts. In the first, we calculate the allowed beat-note frequencies using the expression \eqref{delta-phi} for the round trip phase. The resonant cavity lengths for signal and idler are analyzed in the second part, as well as their relationship with the pump resonance and the wave vector mismatch. 

\subsection{Allowed beat-note frequencies}
\begin{figure}[b]
\includegraphics[scale=0.75]{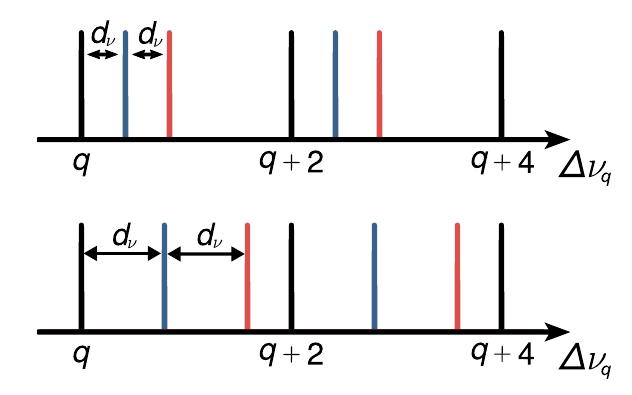}
\caption{Beat-note frequencies (in units of $c/L_{eff}$) calculated from \eqref{beat-note-typei-raw} for $\delta S=0$ (black), $\delta S=1$ (blue) and $\delta S=2$ (red), for a fixed parity of the longitudinal index $q\,$. For the cavity lengths, we considered  $\bar{D}/2R=0.4$ (top) and $\bar{D}/2R=0.9$ (bottom). The shift $d_\nu$ between adjacent combs is given by \eqref{comb-shift}. }
\label{Fig-Gouy-comb}
\end{figure}	

The condition for simultaneous resonance of signal and idler can be written as follows
\begin{eqnarray}
\phi_1&=&2p_1\pi\,,\label{resonance-signal}\\
\phi_2&=&2p_2\pi\,,\qquad p_1,p_2\in\mathbb{Z}\,,\label{resonance-idler}
\end{eqnarray}
which yield
\begin{equation}
\phi_1=\phi_2+2q\pi\,,
\label{acc-phase-constraint}
\end{equation}
where $q=p_1-p_2$. The above relation is required by the double resonance condition and is compatible with the oscillation condition \eqref{equal-detunings} when the the losses for signal and idler are equal. 

We consider a type-I OPO, in which signal and idler have the same polarization. In this case, the difference between the refractive indices for the downconverted fields is not affected by the crystal birefringence, so the frequency dispersion must be taken into account. Following the approach described in Ref.\cite{Martinelli:01}, we expand the refractive indices to the first-order in $\Delta\nu$, obtaining
\begin{eqnarray}
\begin{aligned}
n_1 \approx \bar{n}+\bar{n}^\prime\frac{\Delta \nu}{2}\,,\\
n_2 \approx \bar{n}-\bar{n}^\prime\frac{\Delta \nu}{2}\label{type-I-indices}\,,
\end{aligned}
\end{eqnarray}	 
where  $n^\prime=dn/d\nu$,  and the bars indicate evaluation at $\nu=\nu_0/2$. Substituting \eqref{delta-phi} and \eqref{type-I-indices} in \eqref{acc-phase-constraint}, the resulting approximate expression for the beat-note frequencies is
\begin{equation}
\Delta \nu_q\approx \frac{c}{L_{eff}}\left[q+\frac{2\delta S}{\pi}\arctan\left(\frac{2\bar{z}_0}{2R-\bar{D}}\right)   \right]\,,
\label{beat-note-typei-raw}
\end{equation}
with
\begin{equation}
L_{eff}=\bar
L+l\,\bar{n}^\prime\frac{\nu_0}{2} \left[1+(\bar{S}+1)\frac{\lambda_0}{2\pi\bar{n}^2\bar{z}_0}\right]\,,
\label{FSR-increase-typeI}
\end{equation}
where we recall that the bars refer to the evaluation at frequency degeneracy. The parameters $\bar{S}$ and $\delta S$ are defined as
\begin{eqnarray}
\bar{S}&=&\frac{S_1+S_2}{2}\,,\\
\delta S&=&\frac{S_1-S_2}{2}\,.
\end{eqnarray}

Note that, as in the plane wave case \cite{Eckardt:91,Martinelli:01}, Eq. \eqref{beat-note-typei-raw} yields a comb of beat-note frequencies spaced by $c/L_{eff}\,$, which is the effective free spectral range (FSR) of the OPO.  The effects of diffraction are twofold. First, the average order $\bar{S}$ changes the comb spacing according to Eq. \eqref{FSR-increase-typeI}. However, this dependence is weighted essentially by a factor $\lambda_0/\bar{z}_0$, which is typically small. For a submicron wavelength with a Rayleigh length of a few centimeters, for example, this factor would be smaller than $10^{-4}$.

Second, the combs for different sets of signal/idler transverse modes are shifted from the plane wave solution in proportion to $\delta S$. Considering signal and idler modes with a fixed $\bar{S}$, for example, we see that consecutive values of $\delta S$ yield combs separated by
\begin{equation}
d_\nu=\frac{c}{L_{eff}}\left[\frac{2}{\pi}\arctan\left( \frac{2\bar{z}_0}{2R-\bar{D}}\right) \right]\,,
\label{comb-shift}
\end{equation}
which ranges from zero to a full FSR in the stability limit of the optical cavity ($D=2R$). To illustrate this behavior, we show in Fig.\ref{Fig-Gouy-comb} the result of \eqref{beat-note-typei-raw} for some values of $\delta S$.  We considered a fixed parity of the longitudinal index $q$ due to reasons we detail in Section \ref{Sec-frequency-combs}. Note that for the triangular cavity considered here we have 
\begin{equation}
\frac{2\bar{z}_0}{2R-\bar{D}}=\frac{1}{\sqrt{\frac{}{} 2R/\bar{D}-1}}\,,
\end{equation}
so that the spectral separation between neighboring modes depends only on the ratio $\bar{D}/R$. 

\subsection{Resonant cavity lengths}

	Now we turn to the doubly-resonant cavity lengths. For that we use the resonances of the pump field -- which can be defined regardless of the cavity finesse for the pump -- as a reference. We write the accumulated phase of the pump as
\begin{eqnarray}
\phi_0&=&2p_0\pi+\delta\phi_0\,,\qquad p_0\in\mathbb{Z}\,,\label{pump-phase}
\end{eqnarray}
where $\delta\phi_0$ is the pump detuning at the double resonance.
Now we sum Eqs. \eqref{resonance-signal} and \eqref{resonance-idler}, and subtract from \eqref{pump-phase}. After some algebra, using Eqs. \eqref{delta-phi} and \eqref{detuning-displacement}, we obtain the following expression for the  geometrical distance $\Delta L$ between the double resonance $\{p_1,p_2\}$ and the pump resonance $p_0$
\begin{eqnarray}
\frac{\Delta L}{\lambda_0}=Q+\frac{1}{2\pi}\left(\Delta k l - \Delta\phi_{G}\right)\,,
\label{allowed-pump-detunings}
\end{eqnarray} 
where $Q=p_1+p_2-p_0$ and
\begin{eqnarray}
\Delta\phi_{G}&=&\phi^0_{G}-\phi^1_{G}-\phi^2_{G}\,.\label{Gouy-mismatch}
\end{eqnarray}
The quantity $\Delta \phi_{G}$ is the Gouy phase mismatch, which is the contribution of diffraction to the total phase mismatch.

\begin{figure}[t]
\includegraphics[scale=0.55]{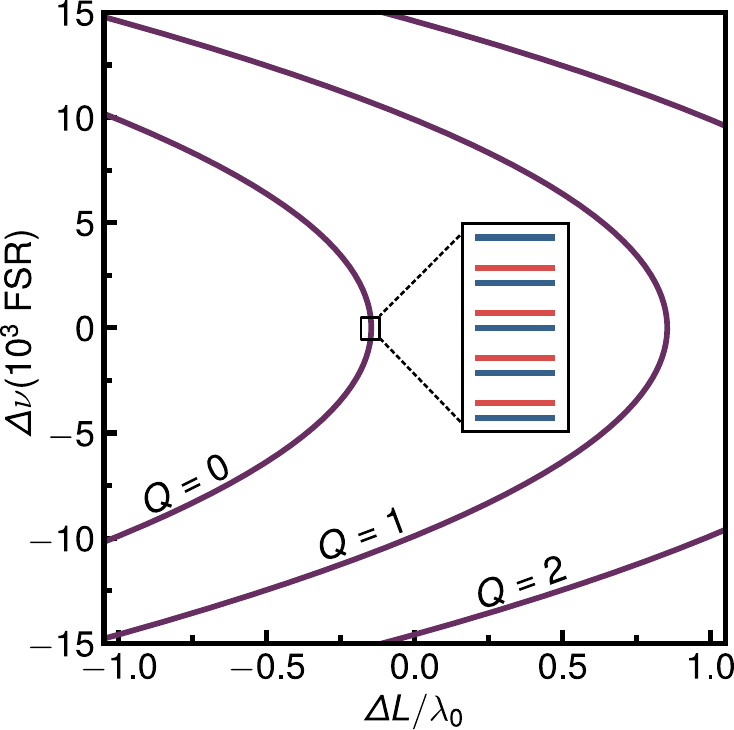}
\caption{Beat-note frequencies versus cavity length in a type-I OPO. The considered downconverted modes are $\{S_1,S_2\}=\{1,1\}$(blue) and $\{2,0\}$(red), with  a second order pump mode $S_0=2$. The inset shows a zoom-in of the degeneracy, with the horizontal lines indicating the allowed frequencies. The crystal parameters are the same as in Appendix \ref{Appendix-bandwidth}, and the cavity length is such that $D/2R\approx0.75$.}
\label{FIG-typeI-beat-notes}
\end{figure}

\section{Spatially structured frequency combs}\label{Sec-frequency-combs}

 In the previous section, we have shown that the OPO spectrum forms a set of equally spaced lines, also known as a frequency comb. In this section, we discuss how a large part of the comb -- a large number of longitudinal modes -- can be produced in an OPO with similar thresholds, and how this depends on the transverse modes involved.
 
	

First of all, let us write the wave-vector mismatch in the more convenient form
\begin{equation}
\Delta k=\frac{2\pi}{\lambda_0}\left[(n_0-\bar{n})-\frac{\bar{n}^\prime \Delta \nu_q^2}{2\nu_0}\right]\,,
\label{wv-mismatch}
\end{equation}	
where we used Eq.\eqref{type-I-indices} for the refractive indices. As $\nu_0\gg \Delta\nu_q \sim c/\bar{L}$ for optical frequencies and $\bar{n}^\prime\,\Delta\nu_q \ll 1$ for smooth dispersion \cite{Lin:91}, the above equation implies that $\Delta k$ only changes appreciably over several free spectral ranges. According to \eqref{beat-note-typei-raw}, this means that $\Delta k$ is a slowly-varying function of the longitudinal index $q$.

Let us consider the change of the resonant cavity lengths with the indices $Q=p_1+p_2-p_0 $ and $q=p_1-p_2$. Substituting substituting \eqref{wv-mismatch} in \eqref{allowed-pump-detunings}, we obtain
\begin{equation}
\frac{\Delta L}{\lambda_0}=Q+\frac{l}{\lambda_0}\left[(n_0-\bar{n})-\frac{\bar{n}^\prime \Delta \nu_q^2}{2\nu_0}\right] -\frac{\Delta\phi_G}{2\pi}\,.
\label{length-beat-note}
\end{equation}
While changing $Q$ by one unit displaces the resonances by $\lambda_0$, changing $q$ yields a much smaller displacement, typically of the order of the cavity FWHM for the downconverted fields (see Appendix \ref{Appendix-bandwidth}). Consequently, if the cavity is locked at the resonance $\{Q,q\}$, the neighboring longitudinal mode $\{Q+1,q+1\}$ is completely off-resonance, but the next one $\{Q,q+2\}$ is only slightly detuned. Therefore, only the modes with a same parity of the index $q$ can have significant overlap between their resonances.


Additionally, the quadratic dependence of $\Delta L$ on $\Delta \nu_q$ implies that the resonances for neighboring longitudinal modes get closer as one approaches the degeneracy. Therefore, the optimal setup for frequency comb generation in an OPO consists of a cavity locked at the resonance for $\Delta \nu_q = 0$ and a crystal phase-matched for degenerate operation. The number of comb modes that are usable in that configuration depends on the tolerance range for the threshold, as we discuss in Appendix \ref{Appendix-bandwidth}.


Regarding the diffraction, this particular condition (frequency degeneracy) occurs at cavity lengths given by \eqref{allowed-pump-detunings}, which depend on the transverse modes via the Gouy phase mismatch. To see how, we expand $\Delta\phi_G$ to the first order in $\Delta \nu_q\,$, obtaining 
\begin{eqnarray}
\Delta\phi_G \! \approx \! \phi_G^0 - 4(\bar{S}\!+\!1)\arctan \! \left(\! \frac{2\bar{z}_0}{2R-\bar{D}}\! \right) 
+\delta S \! \left(\! \frac{l\bar{n}^{\prime}\Delta \nu_q}{2\bar{n}^2\bar{z}_0}\! \right),\nonumber\\
\label{Gouy-mismatch-approx}
\end{eqnarray}
where $\phi^0_G$  is defined in \eqref{Gouy-phase-ring}. Note that the Gouy phase mismatch is mainly affected by $\bar{S}$, since the dimensionless term multiplying $\delta S$ in the above equation is small for smooth dispersion and $\bar{z}_0\sim l$, which apply to typical OPO configurations. This is in clear contrast to the bet-note frequencies \eqref{beat-note-typei-raw}, which depends only on the order difference $\delta S$. This allows multiple frequency combs to be accessed at the same cavity length, corresponding to the combinations of $S_1$ and $S_2$ with $\bar{S}$ fixed.

As a general rule, the OPO produces $\lfloor\bar{S}+1\rfloor$ frequency combs when locked near degeneracy for a given $\bar{S}$ -- the frequencies $\nu_{1,2}$ depend on $|\delta S|\,$, so swapping $S_1$ and $S_2$ does not generate an additional comb. As an example, we show in Fig. \ref{FIG-typeI-beat-notes} the beat-note frequencies versus pump detuning for a second-order pump ($S_0=2$) producing $\{S_1,S_2\}=\{2,0\}/\{0,2\}\,$ and $\{1,1\}\,$, resulting from Eqs. \eqref{allowed-pump-detunings} and \eqref{wv-mismatch}. It should be noted that the solid curves displayed in Fig.\ref{FIG-typeI-beat-notes} are just a guide to the eye, as the allowed beat-note frequencies are discrete.

For a pump with arbitrary spatial structure, the possible transverse modes of the signal/idler are given by the selection rules studied in \cite{Alves2018}. Let us consider the special case of $\Delta S=0$, which encompasses the set of modes with the highest coupling constant. In this case, the nonlinear interaction couples up to $S_0+2$ modes of the downconverted fields in each pair of frequencies of the same comb, mediated by the $S_0+1$ pump modes with order $S_0$.

This connection between the spatial and spectral modes of the OPO is particularly important in the quantum domain, where the nonlinear interaction can be tailored to generate entanglement. For example, Chen \textit{et al.} \cite{Chen:14} demonstrated that a $HG_{00}$ pump mode with multiple frequencies generates large-scale entanglement (60 modes) in the OPO frequency comb. In the same fashion, Liu \textit{et al.} \cite{Liu:16} showed that a monochromatic and multi-structured pump -- a combination of $HG_{20}$, $HG_{02}$ and $HG_{11}$ -- generates quadripartite entanglement in the spatial domain. In both cases, the multipartite entanglement is achieved by tweaking the properties of the pump field.

 In this regard, our results suggest that a single pump structure can be used to entangle multiple sets of modes at once, differing in their frequency spectra and spatial modes. This could be important for generating spatiotemporal graph states, as recently proposed in Ref. \cite{Yang:20}. 
 
Nonetheless, notice that the results up to this point only depend on the orders of the transverse modes involved, which are independent of the mode family. Consequently, the same results apply to Laguerre-Gaussian modes, as long as the crystal astigmatism is negligible. In more extreme astigmatic conditions \cite{Alves2018}, the OPO has no cylindrical symmetry and the Laguerre-Gaussian modes are no longer cavity eigenmodes. The same occurs when the cavity geometry breaks the cylindrical symmetry, as we discuss in the next section.
 
\section{Effect of reflection phase shifts}\label{Sec-reflection-phases}

\begin{figure}
\includegraphics[scale=0.55]{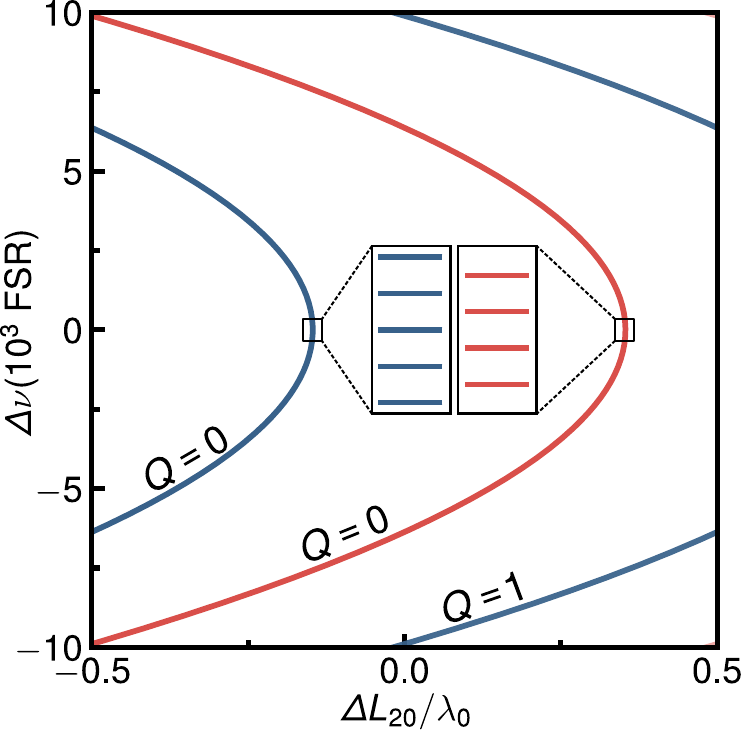}
\caption{Beat-note frequencies versus cavity length with parity phase shifts. The pump mode is a superposition of $HG_{20}$(even) and $HG_{11}$(odd), and the considered interactions are $HG_{20}\rightarrow HG_{10}+HG_{10}$ (blue) and $HG_{11}\rightarrow HG_{10}+HG_{01}$ (red). The insets show a zoom-in of the degeneracy in each curve, with the horizontal lines indicating the allowed frequencies. The system parameters are the same as in Fig.\ref{FIG-typeI-beat-notes}.}
\label{FIG-typeI-beat-notes-sym}
\end{figure}

In this section, we address the reflection phase shifts and their effect on the spatially structured frequency combs. Let us call the phases added by each mirror $\theta_j$, where $j=0,1,2$ labels the pump, signal and idler, respectively. Regardless of the cavity configuration, these phase shifts transform Eq. \eqref{delta-phi} as
\begin{equation}
\phi_j\rightarrow \phi_j+\sum_{i=1}^N \theta_j^{(i)}\,,
\label{transformed-phase}
\end{equation}
where $N$ is the number of reflections in a round-trip. The corresponding changes in the beat-note frequencies and the resonant cavity lengths are
\begin{eqnarray}
\Delta\nu_q &\rightarrow& \Delta\nu_q -\frac{c}{2\pi L_{eff}}\sum^N_{i=1}(\theta_1^{(i)}-\theta_2^{(i)})\,,\label{transformed-beat-notes}\\
\Delta L &\rightarrow& \Delta L + \frac{\lambda_0}{2\pi}\sum^N_{i=1}(\theta_0^{(i)}-\theta_1^{(i)}-\theta_2^{(i)})\,.
\label{transformed-lengths}
\end{eqnarray}

The reflection phase shifts come from two distinct phenomena. One is the boundary condition imposed by the mirror surface, which is usually covered with dielectric coatings. These phases are not necessarily $0$ or $\pi$ -- as one would expect from a metallic surface or the interface between two bulk dielectric media -- and can be engineered to match a given purpose. However, in the particular case of a type-I OPO close to degeneracy, it is reasonable to assume that the coating phases are equal for the downconverted beams, so that \eqref{transformed-beat-notes} leaves the OPO spectrum unchanged.

Second, there is an additional phase shift due to the change of the beam direction. let us assume an arbitrary planar cavity in the $xz$-plane, being $z$ the propagation direction. In this case, a reflection is equivalent to the operation $x\rightarrow -x$, which causes a linearly polarized field with transverse mode $\{m_j,n_j\}$ to be transformed as \cite{Sasada:03}
\begin{equation}
E_j(-x,y,z)\rightarrow (-1)^{\xi_j}E_j(x,y,z)\,,
\label{E-transform}
\end{equation}
where 
\begin{equation}
\xi_j=
\begin{cases}
m_j+1 \qquad x\,\,pol.\,,\\
m_j \qquad \quad y\,\,pol.\,,
\end{cases}
\end{equation}
determines the field parity. The corresponding transformation of the reflection phases is given by
\begin{equation}
\theta_j\rightarrow \theta_j+\xi_j\pi\,,
\label{phase-transformation}
\end{equation}
so that the symmetric and antisymmetric modes acquire a phase difference of $\pi$ in each reflection.

One can readily see that, if $N$ is even,  the transformation \eqref{phase-transformation} leaves \eqref{transformed-beat-notes} and \eqref{transformed-lengths} unchanged, regardless of the transverse modes and polarizations involved. This is the case of all standing-wave cavities, and also the ring cavities with even number of mirrors. For odd $N$, as in a triangular ring cavity, the reflections add an effective phase of  $\pi$ between the symmetric and antisymmetric modes. One can check from \eqref{transformed-beat-notes} that the beat-note combs for $\xi_1=\xi_2$ and $\xi_1=\xi_2+1$ -- with $S_1$ and $S_2$ fixed -- become separated by half a FSR in this case. This is a purely geometrical effect, and hence unlikely to be avoided with coating engineering.

Regarding the resonant cavity lengths, note that the coating phases remain in \eqref{transformed-lengths}, leading to a uniform displacement of the resonant cavity lengths for all sets of transverse modes. For the parity phase shifts, it is worth noting that the nonlinear interaction imposes the following rule \cite{Alves2018}
\begin{equation}
m_0+m_1+m_2=0\,\textrm{(mod 2)} \,.
\end{equation}
As a consequence, $m_0$ and $m_1+m_2$ have the same parity, leading the contribution from the transverse modes to the reflection phases to vanish in \eqref{transformed-lengths}. 

To illustrate the result of the aforementioned constraint, we show in Fig.\ref{FIG-typeI-beat-notes-sym} the calculated beat-note frequencies versus cavity length for a triangular ring cavity with the parity phase shifts. The pump mode is a superposition of $HG_{20}\,$(even) and $HG_{11}\,$(odd), and all beams are y-polarized, so that $\xi_j=m_j$. Note that the $x$ axis is the geometrical distance from the even pump resonance, which is separated from the odd pump resonance by $\lambda_0/2\,$.

\section{Alternative OPO architectures}\label{Appendix-alternative-cavities}

\begin{figure*}
\includegraphics[scale=0.6]{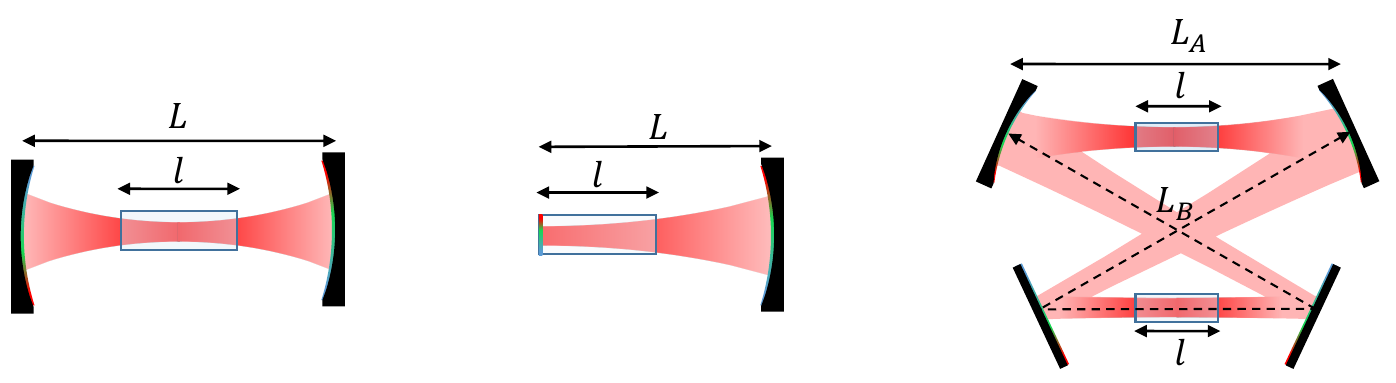}
\caption{Illustration of OPOs designed with (left) a symmetric linear cavity, (center) a semi-monolithic cavity and (right) a symmetric bow-tie cavity.}
\label{Fig-alternative-cavities}
\end{figure*}

The results in Sections \ref{Sec-longitudinal-modes} and \ref{Sec-frequency-combs} were obtained with the expression \eqref{Gouy-phase-ring} for the accumulated Gouy phase, which is for a symmetric ring cavity. In this section, we provide equivalent expressions for other cavity designs, as well as the resulting frequency spectrum in each case.

\subsection{Symmetric linear cavity}

A symmetric linear cavity is formed by to identical concave mirrors, as shown in Fig.\ref{Fig-alternative-cavities}. In this configuration, the expression for the accumulated Gouy phase is the following
\begin{equation}
\phi_G=4(S+1)\arctan\left(\frac{2z_0}{2R-D}\right)\,,
\label{Gouy-phase-linear}
\end{equation}
with $D$ from \eqref{diffraction-length} and 
\begin{equation}
z_0=\frac{1}{2}\sqrt{D(2R-D)}\,.
\label{Rayleigh-length-linear}
\end{equation}
Note that \eqref{Gouy-phase-linear} yields twice the phase of a triangular cavity for the same values of $D$ and $R$. The corresponding expression for the beat-note frequencies is
\begin{equation}
\Delta \nu_q\approx \frac{c}{2L_{eff}}\left[q+\frac{4\delta S}{\pi}\arctan\left(\frac{2\bar{z}_0}{2R-\bar{D}}\right)   \right]\,,
\label{beat-note-linear}
\end{equation}
which presents twice the displacement of the frequency combs due to diffraction, when compared to \eqref{beat-note-typei-raw}. Recall that $L_{eff}$ is given by \eqref{FSR-increase-typeI} and the bars indicate evaluation at $\nu_1=\nu_2=\nu_0/2$.

\subsection{Semi-monolithic cavity}

A semi-monolithic OPO cavity is formed between the coated surface of the nonlinear crystal and one concave mirror, as shown in Fig.\ref{Fig-alternative-cavities}. In this case, the accumulated Gouy phase is given by 
\begin{equation}
\phi_G=2(S+1)\arctan\left(\frac{z_0}{R-D}\right)\,,
\label{Gouy-phase-semimono}
\end{equation}
with $D$ from \eqref{diffraction-length} and
\begin{equation}
z_0=\sqrt{D(R-D)}\,.
\label{Rayleigh-length-semimono}
\end{equation}
The resulting expression for the beat-note frequencies is
\begin{equation}
\Delta \nu_q\approx \frac{c}{2L_{eff}}\left[q+\frac{2\delta S}{\pi}\arctan\left(\frac{\bar{z}_0}{R-\bar{D}}\right)   \right]\,,
\label{beat-note-semimono}
\end{equation}
which is identical to \eqref{beat-note-typei-raw}, up to a relabeling of the cavity and crystal lengths.

\subsection{Symmetric bow-tie cavity}

 A symmetric bow-tie cavity contains two flat mirrors and two identical concave mirrors, as shown in Fig.\ref{Fig-alternative-cavities}. One advantage of this design is that the angles of incidence can be made arbitrarily small, leading to less astigmatism when compared to other ring cavities. Another advantage is that it has two waists, making it favorable for OPOs with two crystals.

 In the scheme shown in Fig.\ref{Fig-alternative-cavities}, the cavity has two segments with geometrical lengths $L_A$ and $L_B$, with a round-trip length given by $L=L_A+L_B$. Considering that the two crystals are identical, the effective diffraction lengths of the segments $A$ and $B$ are 
\begin{eqnarray}
D_A&=&L_A+l\left(\frac{1}{n}-1\right)\,,\\
D_B&=&L_B+l\left(\frac{1}{n}-1\right)\,,
\end{eqnarray}
which are related as
\begin{equation}
D_B=R+\frac{(D_A-R)R^2}{(D_A-R)^2+4z_A^2}\,.
\label{diffraction-lengths-bow-tie}
\end{equation}
The rayleigh lengths $z_A$ and $z_B$ of the segments $A$ and $B$, respectively, are related by
 \begin{eqnarray}
z_B&=&\frac{z_AR^2}{(D_A-R)^2+4z_A^2}\label{Rayleigh-B-bowtie} \,.
\end{eqnarray}

 With the above definitions, the Gouy phase accumulated in a round-trip is given by
\begin{equation}
\phi_G=2(S+1)\arctan\left(\frac{2z_A}{R-D_A}\right)\,,
\label{Gouy-phase-bowtie}
\end{equation}
resulting in the following equation for the beat-note frequencies 
\begin{equation}
\Delta\nu_q\approx-\frac{c}{L_{eff}}\left[q-\frac{2\delta S}{\pi}\arctan\left(\frac{2\bar{z}_A}{R-\bar{D}_A}\right)\right]\,,
\label{beat-note-bowtie}
\end{equation}
with $L_{eff}$ defined in \eqref{FSR-increase-typeI} -- with $z_A$ instead of $z_0$.

\section{Conclusion}\label{conclusion}
 In conclusion, we have studied the spatially structured frequency combs formed by the combination of longitudinal and transverse modes 
of the down-converted fields in an optical parametric oscillator. The diffraction effects are fully considered in the Gaussian beam 
phase structure associated with the curved wavefront. Moreover, the oscillation conditions for frequency combs with higher order 
transverse modes are also established by taking into account the intracavity Gouy phase. This reveals the transverse mode order as 
a key parameter for identifying independent sets of spatially structured frequency combs.
The main effects predicted are
 \begin{enumerate}[(i)]
 \item The beat-note frequencies are displaced from the plane wave solution when the down-converted fields have different mode orders.
 \item The doubly-resonant cavity lengths are displaced from the pump resonance according to the Gouy phase mismatch.
 \end{enumerate}
These effects are crucial for future investigations on the generation of hybrid multipartite entanglement in the quantum optical frequency comb.
Different cavity architectures were considered, which makes our results applicable to a variety of realistic experimental conditions.

\section*{Acknowledgments}

We thank Olivier Pfister for stimulating discussions. This work was supported by Coordena\c c\~{a}o de Aperfei\c coamento de Pessoal de N\'\i vel Superior (CAPES), Funda\c c\~{a}o Carlos Chagas Filho de Amparo \`{a} 
Pesquisa do Estado do Rio de Janeiro (FAPERJ), Instituto Nacional de Ci\^encia e Tecnologia de Informa\c c\~ao Qu\^antica (INCT-IQ) and Conselho Nacional de Desenvolvimento Cient\'{\i}fico e Tecnol\'ogico (CNPq).

\appendix

\section{Tolerance to phase-mismatch and detuning}\label{Appendix-bandwidth}

As we mentioned in Section \ref{Sec-frequency-combs}, the modes of the OPO frequency comb resonate at separate cavity lengths and correspond to different values of the wave-vector mismatch. In this appendix, we address the number of modes that can be produced with thresholds close to minimum -- and hence squeezing levels close to maximum -- in an OPO near degeneracy, considering the phase-mismatch and the detuning as the limiting factors. 

First, let us consider the thin crystal limit and a ring OPO, in which the diffraction effect on the coupling strength can be neglected. In this case, we have 
\begin{equation}
g\approx g_0.\textrm{sinc}(\Delta k l/2)\,,
\end{equation}
where $g_0=g|_{\Delta k=0}$ and $\textrm{sinc}(x)=\sin(x)/x$.
If the crystal is phase-matched for $q=0$ and locked at the perfect resonance for that mode, the threshold for $q=2N$ is $I_N=\mu I_0$, where $I_0$ is the threshold for $q=0$ and 
\begin{equation}
\mu=\frac{1+F^2\delta\phi_N^2/\pi^2}{\textrm{sinc}^2(\Delta k_N l/2)}\,.
\label{threshold}
\end{equation}
In the above equation, $F$ is the cavity finesse, $\delta \phi_N=\delta \phi_{1,2}|_{q=2N}$ is the common detuning of the signal and idler fields in the Nth longitudinal mode and $\Delta k_N$ is the corresponding value of the wave-vector mismatch.  We have also assumed the thin crystal limit, in which the diffraction effect on the coupling strength can be neglected. 

Owing to the frequency dispersion of the refractive indices, both the wave-vector mismatch and the detuning increase as one moves away from the degeneracy. A given tolerance for the corresponding increase in the threshold constrains the OPO spectrum to a bandwidth around the frequency degeneracy, which we call threshold bandwidth. In the following, we consider a maximum $1\%$ increase above the optimal threshold ($\mu=1.01$), following Ref. \cite{Wang:14}.

The wave-vector mismatch for the Nth mode is obtained by substituting \eqref{beat-note-typei-raw} in \eqref{wv-mismatch}. Considering $\delta S=0$, for simplicity, we obtain
\begin{equation}
\frac{\Delta k_N l}{2}=-4\pi\Gamma N^2\,,
\end{equation}
where
\begin{equation}
\Gamma=\frac{l\lambda_0}{L_{eff}^2}\frac{\bar{n}^\prime\nu_0}{2}\,.
\label{Gamma}
\end{equation}
Note that $\Gamma$ is a dimensionless parameter that depends on both the crystal and cavity characteristics. Consider the example of a LBO crystal with the downconverted fields polarized along the ordinary axis, in which case the Sellmeier equations \cite{Lin:91} yield $\bar{n}^\prime\nu_0/2\approx10^{-2}$ for $\lambda_0=532nm$. Assuming the typical values $L_{eff}=10cm$, $l=1cm$ for the cavity and crystal lengths, we obtain $\Gamma\approx10^{-8}$.

As for the detuning, we need to calculate the geometrical displacement between the resonances for $q=0$ and $q=2N$. From Eqs.\eqref{beat-note-typei-raw}, \eqref{allowed-pump-detunings} and \eqref{wv-mismatch}, we obtain 
\begin{equation}
\begin{split}
d_N=&|\Delta L(Q,q=2N)-\Delta L(Q,q=0)|\\
=&\, 4 \lambda_0 \Gamma N^2\,,
\end{split}
\label{resonance-distance}
\end{equation}
with $\Gamma$ from \eqref{Gamma}. According to \eqref{detuning-displacement}, this results in a detuning given by
\begin{equation}
\begin{split}
 \delta \phi_N \approx&\frac{\pi d_N}{\lambda_0} \\
=& \,4\pi\Gamma N^2\,,
 \end{split}
 \label{detuning-comb}
 \end{equation} 
where we considered $\lambda_1=\lambda_2=2\lambda_0$ as an approximation. Substituting \eqref{resonance-distance} and \eqref{detuning-comb} in \eqref{threshold}, we obtain the following expression for $\mu$
\begin{equation}
\mu=\frac{1+(4F\Gamma N^2)^2}{\textrm{sinc}^2(4\pi\Gamma N^2)}\,.
\end{equation}
Taking $\mu=1.01$ with $F=100$, for example, we obtain a total of  $N_{BW}\approx 300$ longitudinal modes within the threshold bandwidth. This amounts to a maximum displacement from resonance of $\sim 5\%$ of the FWHM for the downconverted fields.

The threshold bandwidth can be expanded by decreasing either $\Gamma$ or $F$. Smaller values for $\Gamma$ can be obtained by using a larger cavity or a crystal with broader frequency dispersion, as in Ref.\cite{Wang:14}. On the other hand, a smaller finesse is obtained by increasing the cavity losses, and hence the FWHM of the signal/idler resonances. A higher FWHM mitigates the effect of dispersion, making it more difficult for neighboring resonances to lose the overlap.

\bibliographystyle{apsrev4-2} 
\bibliography{references} 

\end{document}